\documentclass[noshowpacs,amsmath,
twocolumn,
superscriptaddress,
8pt,
aps,prb]{revtex4-1}  % this article uses the Revtex 4-1 format
\usepackage{setspace}
\usepackage{amsmath}
\usepackage{graphicx}
\usepackage[nearskip,margin = 0pt]{subfig}
\usepackage{verbatim}  % for commenting
\usepackage{amsfonts}
\usepackage{amssymb}
\usepackage{epstopdf}
\usepackage{xcolor}
\DeclareGraphicsExtensions{.pdf,.eps,.png,.jpg,.mps}

\begin{document}

\title{Broadband dispersion engineered microresonator on-a-chip}

\author{Ki Youl Yang$^1$, Katja Beha$^2$, Daniel C. Cole$^2$, Xu Yi$^1$, Pascal Del'Haye$^2$,  Hansuek Lee$^1$, Jiang Li$^1$, Dong Yoon Oh$^1$,\\
 Scott A. Diddams$^2$, Scott B. Papp$^2$, and Kerry J. Vahala$^1$\\
$^1$T. J. Watson Laboratory of Applied Physics, California Institute of Technology, Pasadena, California 91125, USA.\\
$^2$Time and Frequency Division, National Institute of Standards and Technology, Boulder, Colorado, 80305 USA.}

\maketitle

%%%%%%%%%%%%%%%%%%%%%%%%%%%%%%%%%%%%%%%%%%%%%%%%%%%%%%%%%%%%%%%%%%%%%%%%%%%%
%%%%%%%%%%%%%%%%%%%%%%%%%%%%%%%%%%%%%%%%%%%%%%%%%%%%%%%%%%%%%%%%%%%%%%%%%%%%
%%%%%%%%%%%%%%%%%%%%%%%% Intro section %%%%%%%%%%%%%%%%%%%%%%%
%%%%%%%%%%%%%%%%%%%%%%%%%%%%%%%%%%%%%%%%%%%%%%%%%%%%%%%%%%%%%%%%%%%%%%%%%%%%

\textbf{Control of dispersion in fibre optical waveguides is of critical importance to optical fibre communications systems\cite{Agrawal_2007,Ainslie_1986} and more recently for continuum generation from the ultraviolet to the mid-infrared\cite{ranka_2000_visible,Reeves_2003_transformation,Dudley_2006_Review,petersen_2014_mid}.  The wavelength at which the group velocity dispersion crosses zero can be set by varying fibre core diameter or index step\cite{Cohen_1979,White_1979,tsuchiya1979dispersion,Petermann_1983,Ainslie_1986}. Moreover, sophisticated methods to manipulate higher-order dispersion so as shape and even flatten dispersion over wide bandwidths are possible using multi-cladding fibre\cite{kawa_1974characteristics, Miya_1981fabrication,Cohen_1982QCfiber,Jang_1982Exp,bhagavatula_1983, Monerie_1982propagation,QC_design_1984}.  Here we introduce design and fabrication techniques that allow analogous dispersion control in chip-integrated optical microresonators, and thereby demonstrate higher-order, wide-bandwidth dispersion control over an octave of spectrum.  Importantly, the fabrication method we employ for dispersion control simultaneously permits optical Q factors above 100 million, which is critical for efficient operation of nonlinear optical oscillators. Dispersion control in high Q systems has taken on greater importance in recent years with increased interest in chip-integrable optical frequency combs\cite{Pascal_comb_Nature,Pascal_2009NP,Kippendberg_Diddams_review,Pascal_2011_octave,Okawachi_2011_SiN,Herr_2012_NP,
Jiang_2012_PRL,Coen_2013_modeling,Chembo_2013_PRA,Lamont_2013_SiNModel,
Pascal_2014_PRL,Herr_2014_Solition,Herr_2014_PRL_mode,Papp_2014_Optica,
Pascal_2014_Arx,Grudinin_2014micro,Okawachi_2014_SiN,Brasch_2014_SiN,Xue_2014_dark,Jost_2014_microwave}.}\\
\indent
%%%%%%%%%%%%%%%%%%%%%%%%%%%%%%%%%%%%%%%%%%%%%%%%%%%%%%%%%%%%%%%%%%%%%%%%%%%%
%%%%%%%%%%%%%%%%%%%%%%%%%%%%%%%%%%%%%%%%%%%%%%%%%%%%%%%%%%%%%%%%%%%%%%%%%%%%
%%%%%%%%%%%%%%%%%%%%%%%% Main Text %%%%%%%%%%%%%%%%%%%%%%%
%%%%%%%%%%%%%%%%%%%%%%%%%%%%%%%%%%%%%%%%%%%%%%%%%%%%%%%%%%%%%%%%%%%%%%%%%%%%

High-Q microresonators, through cavity enhancement of an input field, enable many important optical devices and functions\cite{Vahala_2003_Nature}. These applications include frequency comb generation\cite{Pascal_comb_Nature,Pascal_2009NP,Kippendberg_Diddams_review,Pascal_2011_octave,Okawachi_2011_SiN,Herr_2012_NP,
Jiang_2012_PRL,Coen_2013_modeling,Chembo_2013_PRA,Lamont_2013_SiNModel,
Pascal_2014_PRL,Herr_2014_Solition,Herr_2014_PRL_mode,Papp_2014_Optica,
Pascal_2014_Arx,Grudinin_2014micro,Okawachi_2014_SiN,Brasch_2014_SiN,Xue_2014_dark,Jost_2014_microwave}, pulse sources\cite{Peccianti_2012_NC,Pasquazi_2012_OE}, cascaded Raman lasers\cite{Spillane_2002_Raman,Rong_2005_Raman}, stimulated Brillouin lasers\cite{Hansuek_2012_NP,Buttner_2014_SBS} and harmonic generation \cite{Carmon_2007_visible,Levy_2011_OE}. In many systems the control of modal dispersion is helpful or essential. This is particularly important in parametric oscillators\cite{kippenberg_2004PRL,grudinin_maleki_2009} where the local dispersion must be anomalous and in frequency microcombs where both the sign of the dispersion as well as its spectral shape are critical for comb operation\cite{Kippendberg_Diddams_review,Pascal_comb_Nature}. Spectral bandwidth and coherent operation are strongly influenced by dispersion, and there has been remarkable progress in these areas, including sub-comb synchronization\cite{Pascal_2014_PRL,Papp_2014_Optica,Jiang_2012_PRL,Pascal_2014_Arx}, soliton generation\cite{Herr_2014_Solition,Herr_2014_PRL_mode,Brasch_2014_SiN,Xue_2014_dark,Jost_2014_microwave}, and dispersive wave formation\cite{Lamont_2013_SiNModel,Okawachi_2014_SiN,Coen_2013_modeling,Brasch_2014_SiN}.\\
\indent 
There have been multiple approaches to engineer dispersion in microresonators. The control of waveguide width and height in Si$_{3}$N$_{4}$, diamond, MgF$_{2}$,  and CaF$_{2}$ resonators alters geometric dispersion\cite{Okawachi_2011_SiN,Okawachi_2014_SiN,Grudinin_2014micro,Hausmann_2014_diamond}. HfO$_{2}$-coated Si$_{3}$N$_{4}$ and oxidized Si resonators also provides dispersion control\cite{Riemensberger_2012_OE,Jiang_2014_Sidispersion}.
\begin{figure*}[t!]
  \begin{centering}
  \includegraphics[width=17 cm]{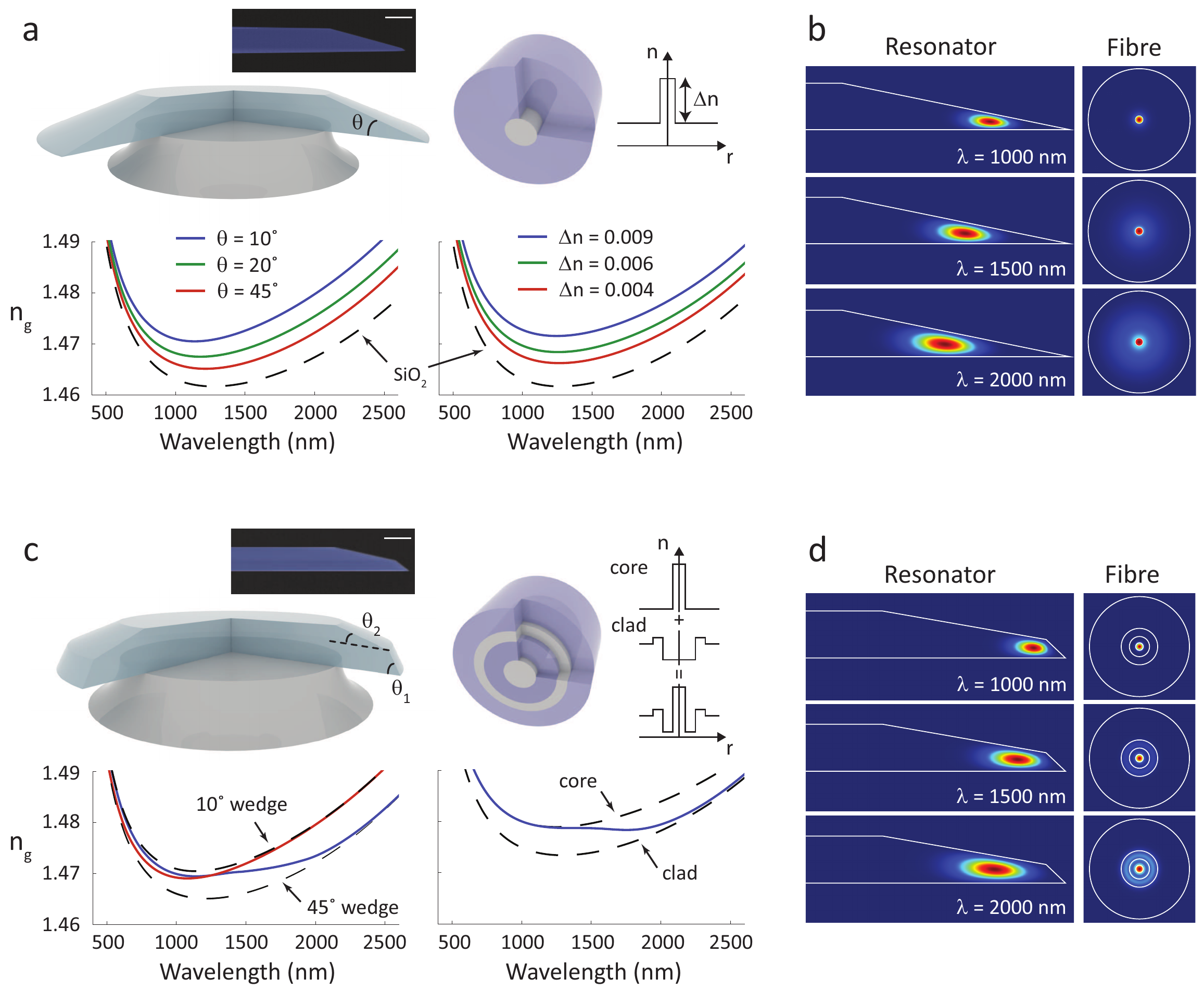}
  \captionsetup{singlelinecheck=no, justification = RaggedRight}
  \caption{\textbf{Fibre-inspired cavity dispersion design.} (\textbf{a}) Upper panels depict single-wedge resonator and single-core fibre; and lower panels give the corresponding group-index spectra of the fundamental modes in the single-wedge disk (left panel) and single-clad fibre (right panel). Here, ${\theta}$ is the wedge angle of the resonator, and ${\Delta} n$ is core-cladding refractive index difference. The dashed curves give the bulk group index of silica. Calculations assume a resonator diameter of 3 mm and fibre core diameter of 11 ${\mu}m$. Micrograph of resonator cross section and fibre refractive index profile are provided as insets. White scale bar is 10 ${\mu}m$. (\textbf{b}) Finite element simulation of the fundamental mode in a single-wedge disk (left panel) and single-clad fibre (right panel) at the wavelengths 1000, 1500, and 2000 nm. The index contrast and core diameter in the fibre mode calculation has been set to ${\Delta}$n = 0.0028 and 4.5 ${\mu}m$ so as to make the variation in mode profile with wavelength more readily observable.  (\textbf{c}) As in panel \textbf{a}, but for double-wedge disk and multi-clad fibre. The dashed lines in the group index spectra give either the single wedge results (10$^{\circ}$ and 45$^{\circ}$ cases) or the single core and clad cases for the fibre. The blue curve in the lower right panel gives the multi-clad fibre case. The blue and red curves in the lower left panel give the multi-wedge resonator in cases where the angle ordering is red: 45$^{\circ}$ (outer)$\to$10$^{\circ}$ (inner) and blue: 10$^{\circ}$ (outer)$\to$45$^{\circ}$ (inner). (\textbf{d}) Same as panel \textbf{b}, but for a double-wedge resonator and multi-clad fibre. The core diameter in the multi-clad fibre plots are set to 4.5 ${\mu}m$ to make the variation in mode profile with wavelength more readily observable.}
\label{fig1}
\end{centering}
\end{figure*}
\begin{figure*}[t!]
  \begin{centering}
  \includegraphics[width=18cm]{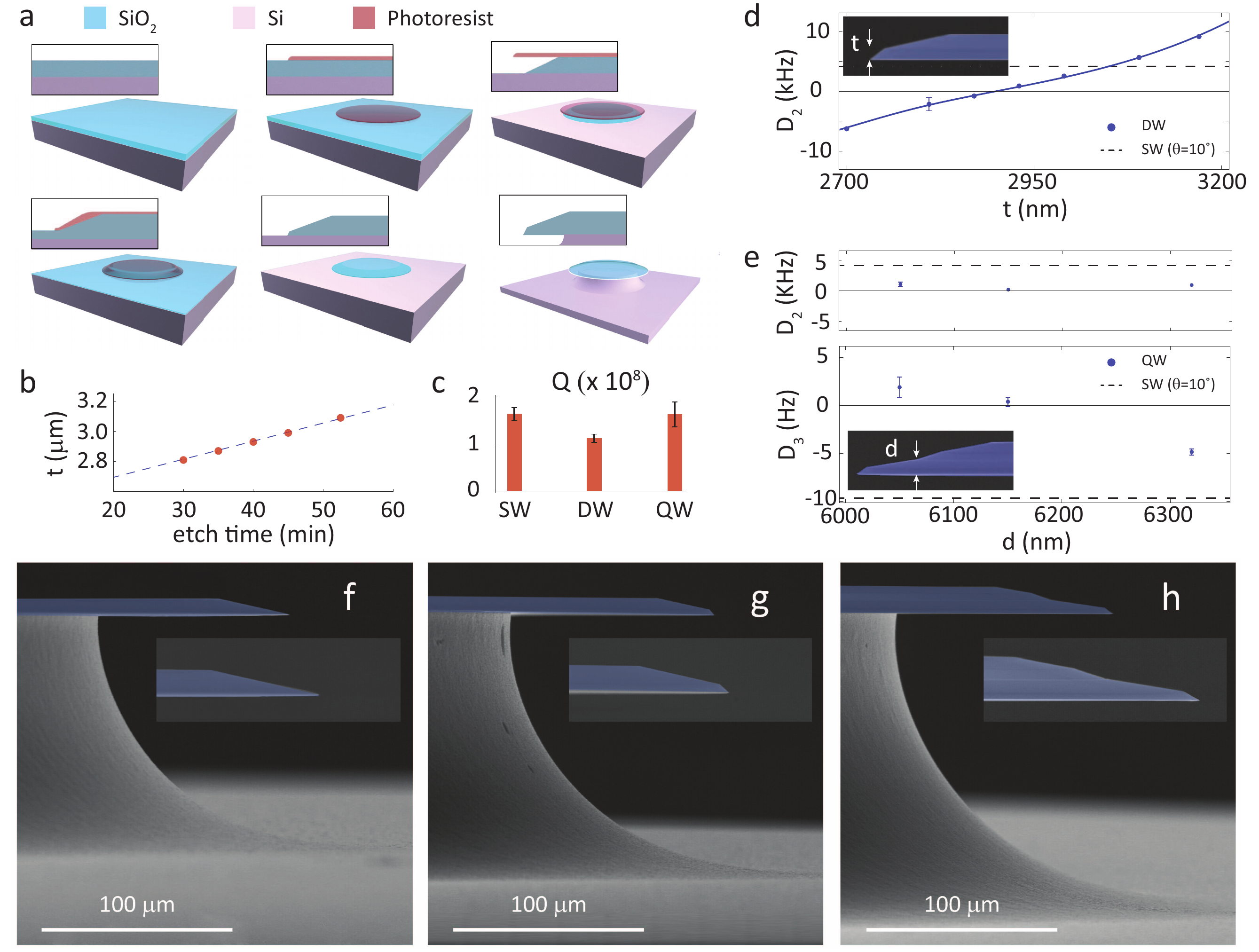}
  \captionsetup{singlelinecheck=no, justification = RaggedRight}
  \caption{\textbf{Microfabrication process flow and side-view micrographs.} (\textbf{a}) Fabrication process flow steps required to fabricate a double-wedge resonator. (\textbf{b}) Measured wedge height (``t'', see inset of \textbf{d}) control as a function of etch time. (\textbf{c}) Measured Q factor of TM fundamental mode in single, double, and quadruple wedge disks. (\textbf{d}) ${D}_{2}$ (=$\Delta$FSR per mode) as a function of the outer wedge height (``t'', see inset) of double-wedge disk. (\textbf{e}) ${D}_{3}$ as a function of the third wedge height (``d'', see inset) of quadruple-wedge disk. Outermost wedge height is designed to make ${D}_{2}$ within 0 - 1 kHz (see upper panel). (\textbf{f-h}) Scanning electron micrographs of side-views for (\textbf{f}) single,  (\textbf{g}) double and  (\textbf{h}) quadruple wedge resonators.}
\label{fig2}
\end{centering}
\end{figure*}
\begin{figure*}[t!]
  \begin{centering}
  \includegraphics[width=17cm]{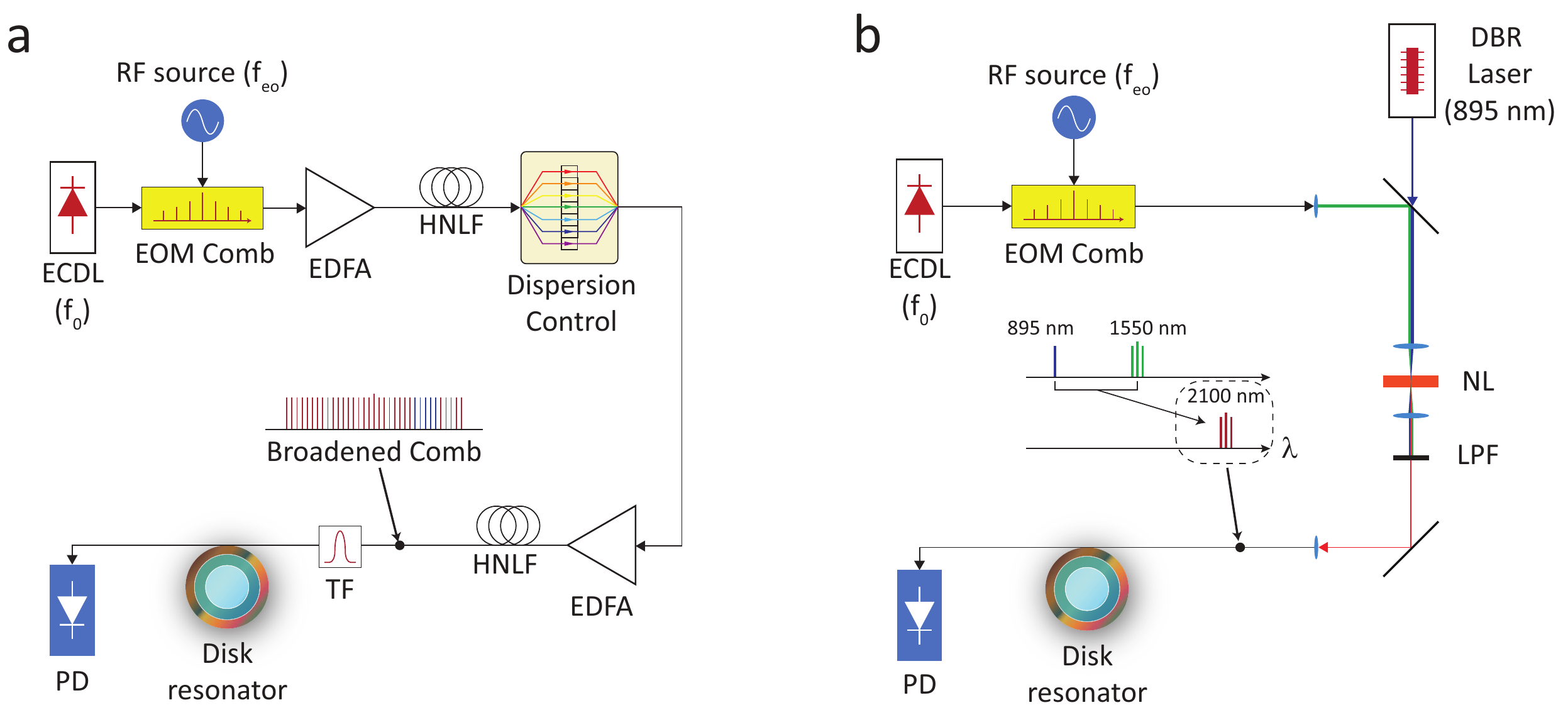}
   \captionsetup{singlelinecheck=no, justification = RaggedRight}
    \caption{\textbf{Dispersion measurement setup.} (\textbf{a}) Broadband (1400$-$1700 nm) FSR measurement setup using externally broadened EOM comb. ECDL: External cavity diode laser, EDFA: Erbium doped fibre amplifier, HNLF: Highly nonlinear fibre, TF: Tunable bandpass filter, PD: Photodetector. (\textbf{b}) FSR measurement setup at 2100 nm using difference frequency generation between a 895 nm distributed Bragg reflector (DBR) laser and 1550 nm EOM comb. NL: Nonlinear crystal, LPF: Lowpass optical filter  }
\label{fig3}
\end{centering}
\end{figure*}
This paper studies dispersion control in ultra-high-Q (UHQ) silica disks using a method that is inspired by dispersion engineering in optical fibres. By creating multi-wedge disks and precisely controlling their geometry during microfabrication, we have devised technique that lithographically controls higher-order dispersion over an octave of bandwidth. After introducing the approach and microfabrication method, resonator dispersion is measured at various bands from 960 nm to 2100 nm and compared with modeling to confirm dispersion control. \\
\\
\textbf{Results}\\
\textbf{Dispersion design principle.} Optical fibre designers use multiple cladding layers to control both the magnitude, sign and spectral profile of the combined material and waveguide dispersion\cite{Miya_1981fabrication,Jang_1982Exp,Cohen_1982QCfiber,bhagavatula_1983}. We devised a method for dispersion control in disk resonators that is analogous to these methods. Fig. 1a compares the dispersion of a single-wedge disk with that of a single-core optical fibre. Details on fabrication and basic properties of the single-wedge disk are contained in \mbox{ref [\!\!\citenum{Hansuek_2012_NP}]}. Briefly, the device is fabricated from thermal silica on silicon using lithography combined with wet and dry etching. In terms of dispersion control a key parameter is the wedge angle ${\theta}$. The group index spectra of three resonators that differ only in their wedge angles are shown in the figure. The group index of bulk silica is also provided for comparison. In effect, the wedge angle introduces a component of normal dispersion that is stronger for smaller wedge angles\cite{Hansuek_2012_NP,JiangLi_OE}. By comparison, in the single-clad fibre\cite{tsuchiya1979dispersion}, the group index is varied (for fixed core diameter) through control of ${\Delta}n$. \\
\indent
Fig. 1b shows the calculated optical mode in the single-wedge disk and the optical fibre to further understand the mechanism of dispersion control in each system. In the resonator, longer wavelength modes see their centroid of motion around the resonator shift inwards (i.e., smaller radii) so that the effective optical path is smaller. This is equivalent to normal dispersion\cite{Hansuek_2012_NP,JiangLi_OE}. In the fibre, longer wavelength modes have a greater spatial overlap with the lower-index cladding. Greater contrast between the core and cladding indices therefore strengthens this normal contribution to dispersion. \\
\indent
This similarity between the resonator and fibre cases suggests a method to engineer dispersion in the resonator that is illustrated in the Fig. 1c. In the right-hand panel, the case of a multi-clad fibre is considered in relation to a resonator featuring multiple wedge angles. The multi-clad fibre can be understood by analyzing the dispersion in short and long-wavelength cases\cite{Cohen_1982QCfiber}. Short wavelength modes are confined primarily by the core region and experience a group index that is similar to a corresponding single core fibre. Longer wavelength modes are primarily confined by the secondary cladding layer. The corresponding spectral dependence of the group index can be understood as a transition between these two extreme cases (see spectrum for multi-clad fibre in Fig. 1c). The wavelength at which the transition occurs can be correspondingly engineered\cite{kawa_1974characteristics, Monerie_1982propagation,QC_design_1984}. For example, by placing the outer, higher index region closer to the inner core, the transition will occur at shorter wavelengths. \\
\indent
Using analogous reasoning for the resonator, the case of a double-wedge resonator is considered in the left-hand panel of Fig. 1c. In this case, the shorter wavelength modes experience the outer, larger-angle wedge (${\theta}_{1}$=45$^{\circ}$) while the longer wavelength modes experience the inner, smaller wedge angle (${\theta}_{2}$=10$^{\circ}$). The spectral dependence of the group index thereby transitions between these two extreme cases. By controlling the specific wedge angles the long and short wavelength limiting behavior is controlled. Likewise, in analogy with the multi-clad fibre, the location of the wedge-angle transition allows control of the wavelength band over which the group index transitions between these two limits. As shown below, this concept can be extended to the design the group index of a quadruple-wedge disk which provides more flexibility in dispersion control over a wider range of wavelengths.\\
\indent
\\
\textbf{Device structure and fabrication process.} The multi-wedge disk process is an extension of the techniques developed for the original single-wedge disk resonator\cite{Hansuek_2012_NP}. Fabrication begins with thermally grown SiO${_{2}}$ on Si, which is then processed with photolithography and wet etching into single-wedge shaped disks (see top row in Fig. 2a) in which the wedge angle can be controlled from 8${^{\circ}}$ to 55${^{\circ}}$ (see Methods). A double-wedge (see lower row in Fig. 2a) is created through another cycle of thermal oxidation, photolithography and wet etching. During this second cycle, the angle of the additional wedge is adjusted as before and its height is controlled through the duration of etching. Calibrations have shown that the height increases at a rate of ${\sim}$ 10 nm/min (cf. Fig. 2b), resulting in the following spectral rates: FSR ${\sim}$ 400 kHz/min and ${\Delta}$FSR ${\sim}$ 300 Hz/min (cf. Fig. 2d). The process can be repeated to add additional wedges. A four wedge structure is shown in Fig. 2h. Because of the added complexity of the process, the optical Q factors of the multi-wedge structures are lower than for the single-wedge devices. Nonetheless, Q factors of 1.63, 1.12, 1.62 $\times$ 10$^{8}$ for single-, double-, and quadruple-wedge, respectively were achieved at 1550 nm. 
\begin{figure}[t!]
  \begin{centering}
  \includegraphics[width=8.5cm]{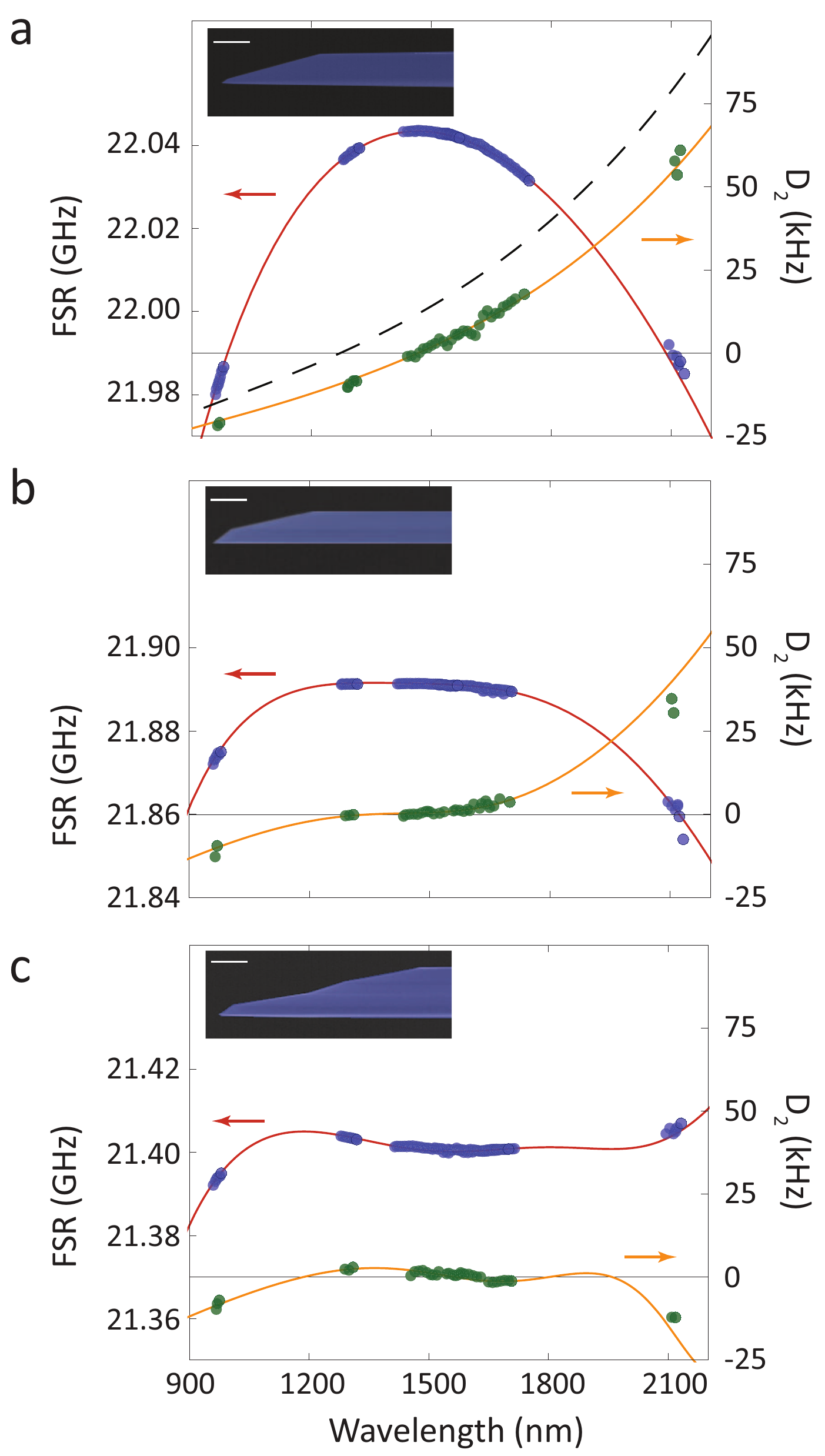}
   \captionsetup{singlelinecheck=no, justification = RaggedRight}
   \caption{\textbf{Dispersion measurements for single-, double-, and quadruple-wedge disks.} FSR, and ${D}_{2}$ for (\textbf{a}) single-, (\textbf{b}) double-, (\textbf{c}) quadruple-wedge disk. Blue and green data points are measured FSR and  ${D}_{2}$, respectively; red and orange lines are numerically calculated FSR and ${D}_{2}$, respectively. The cavity geometry is directly imported from the scanning electron micrograph image (inset, bar: 10 ${\mu}m$) for finite element simulations.}
\label{fig4}
\end{centering}
\end{figure}
\\
\indent
\\
\textbf{Dispersion characterization.} To characterize dispersion in the resonator we measure $FSR_{\mu}$ (${\mu}$ is mode index) in several wavelength bands and calculate ${D}_{2}(\mu)=\partial FSR_{\mu'}/\partial \mu'|_{\mu'=\mu}$ which is related to the GVD parameter through the following expression:
\begin{eqnarray}
{\beta}_{2}(\mu) = \frac{\partial^2 \beta}{\partial \omega^2}\bigg|_{\omega=\omega_{\mu}}\approx-\frac{1}{4\pi^2 R}\cdot\frac{{D}_{2}(\mu)}{FSR_{\mu}^3}
\end{eqnarray}
where R is a resonator radius\cite{Jiang_2014_Sidispersion,Pascal_2009NP}. As a preliminary test of ${D}_{2}$ (at 1550 nm) control in double-wedge design, a series of the outer wedge heights (``t", see inset) were fabricated with different wet etch times. A plot of the measured ${D}_{2}$ versus outer wedge height is provided in Fig. 2d, and the solid line is calculated using a finite element solver. ${D}_{2}$ increases as a function of etch duration at the average rate of 300 Hz/ min, and the measured ${D}_{2}$ has  a standard deviation of 495 Hz from the finite-element-simulation resulting from the available precision in the atomic force microscope and the scanning electron microscope measurement of the resonator cross-section. \\
\indent
An additional interior wedge adds geometric dispersion (normal dispersion) in the spectral range that optical modes experience the new wedge angle. Fig. 2e shows measurements of ${D}_{2}$ and ${D}_{3}$ in such a structure plotted versus the parameter ``d" (see inset in Fig. 2e). Here, we define higher order dispersion term as
\begin{eqnarray}
{f}_{\mu} ={f}_{0}+D_{1}\cdot {\mu}+\frac{D_{2}}{2!}{\mu}^{2}+\frac{D_{3}}{3!}{\mu}^{3}+\cdots
\end{eqnarray}
where ${f}_{0}$ is the frequency at which the dispersion is measured and ${D}_{1}$, ${D}_{2}$, ${D}_{3}$ correspond to FSR, ${\Delta}$FSR per mode, and the third order dispersion parameter\cite{Herr_2014_Solition,Herr_2014_PRL_mode}. As shown in measurements below, the slope of dispersion curve at 1550 nm gets flatter (${D}_{3}\approx0$) as ``d" decreases and eventually becomes inverse (${D}_{3}>0$) to the slope of material dispersion curve. In addition, the outer wedge height is also designed to keep ${D}_{2}$ consistent within a range of $0 - 1$ kHz. \\
\indent
To measure the FSR quickly and over a broad range of wavelengths, we have modified a method reported in \mbox{ref [\!\!\citenum{JiangLi_OE}]}. The measurement setup is shown in Fig. 3a and incorporates a frequency comb generated using electro-optical modulation (EOM)  in conjunction with optical broadening using high nonlinearity fibre (HNLF). The seed laser for the EOM comb was a tunable laser that was scanned continuously during the measurement.  Further details on this EOM comb are provided in \mbox{ref [\!\!\citenum{Katja_FIO}]}. The comb bandwidth was approximately 300 nm centered around 1550 nm and the line spacing was tunable and set to coincide approximately with the resonator FSR. A relatively narrow band of comb frequencies was coupled to the resonator using a tunable optical filter and the transmitted optical signals were measured using a photodetector. A measurement proceeded by first setting the tunable filter to a desired wavelength.  The optical filter had a 3-dB-bandwidth of 3 nm so approximately 15 comb teeth would be launched to the resonator input port (the EOM comb line spacing was approximately 22 GHz). The photo detected signal was then observed as a time trace on an oscilloscope that was synchronized with the scanning seed laser. If the microwave frequency used to establish the EOM comb line spacing was set to exactly coincide with the FSR of the resonator at the selected wavelength, then a single transmission minimum would appear since the filtered comb teeth are all coupled to the neighboring resonances of the resonator. By slightly tuning the comb drive frequency away from the FSR value, the single transmission minimum would break up into many separate peaks since the comb teeth would now achieve resonance at slightly different points in the laser scan. The resolution in determination of the FSR by this method was estimated to be about 100 kHz. It is also important to note that the method provides an FSR value that is spectrally averaged since multiple comb teeth participate in the process. This fact tends to smooth the data with respect to effects like avoided-crossings of mode families. In addition to measurements in the 1500 nm band, the EOM comb was also shifted in wavelength to 2.1 ${\mu}m$ by difference frequency generation using a 895 nm pump laser. Also, measurements of FSR using the technique reported in ref \mbox{[\!\!\citenum{JiangLi_OE}]} were conducted at 980 nm and 1300 nm. Collectively, these measurements provided a very good picture of how dispersion could be modified through judicious cavity design. \\
\indent
Fig. 4a-c show measured FSR, ${D}_{2}$, and the results of simulations for the device structures provided in the insets. As has been noted in \mbox{ref [\!\!\citenum{JiangLi_OE}]}, a single wedge contributes geometrical dispersion that is both normal and approximately spectrally flat. Also, smaller wedge angles increase the strength of this contribution. In figure 4a, the zero dispersion wavelength (${\lambda}_{ZDW}$), has been shifted by this effect to 1.5 ${\mu}m$\cite{Hansuek_2012_NP,JiangLi_OE} (material dispersion has a ${\lambda}_{ZDW}$ near 1.3 ${\mu}m$).\\
\indent
In the double wedge resonator (Fig. 4b) there is a gradual onset of the geometrical dispersion from shorter to longer wavelengths as a result of a steep outer wedge (low normal dispersion) to the shallow interior wedge (larger normal dispersion). The overall effect is to provide a flattening of the dependence of FSR on wavelength over a broad range of wavelengths. The dispersion in the transition is therefore reduced. The addition of additional wedges can be used to extend control to longer wavelengths. In the quadruple-wedge-design the additional interior wedges provide control out to 2 ${\mu}m$ (Fig. 4c). \\
\indent
\\
\textbf{Conclusion}\\
This study has proposed and tested a resonator dispersion control approach that is inspired by dispersion control methods in multi-cladding optical fibre. This approach provides the first method of controlling resonator dispersion over a broad range of wavelengths (equivalent to having control over high-order dispersion). The approach builds on the high-Q silica-on-silicon wedge-resonator fabrication by adding additional wedge angles to the resonator. The position of each wedge and its angle determine the wavelength location and strength of a corresponding contribution to normal dispersion. As a result, the approach is highly intuitive and when combined with numerical modeling provides a well controlled way to engineer dispersion over broad wavelength bands. Dispersion engineered high-Q resonators, as described here, will be useful in nonlinear resonator devices such as frequency microcombs. Moreover, the methods can also be applied in control of dispersion in nonlinear waveguide structures.  \\
\\
\textbf{Methods}\\
\textbf{Device fabrication.} The fabrication process was based on the previously described method for the chemically etched wedge-resonator\cite{Hansuek_2012_NP}. Disk resonators were fabricated on (100) prime-grade float-zone silicon wafers. The first SiO$_{2}$ layer was thermally grown at 1000$^{\circ}$C with a thickness in the range of 7$-$9 ${\mu}$m, and photoresist was patterned using a KarlSuss MA-6 aligner on the oxide layer. 
\begin{figure}[h]
  \begin{centering}
  \includegraphics[width=8cm]{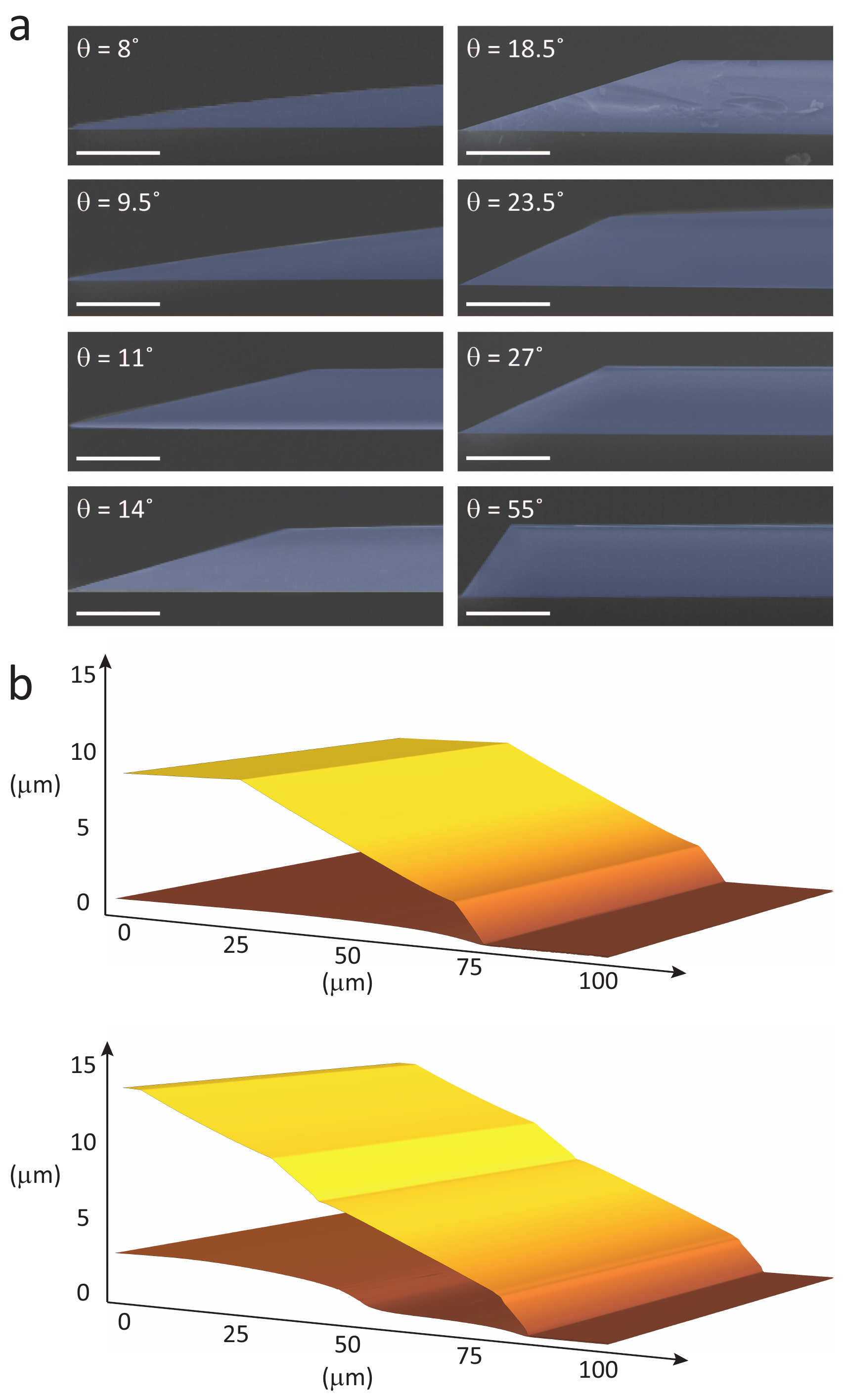}
   \captionsetup{singlelinecheck=no, justification = RaggedRight}
    \caption{\textbf{Wedge angle control and atomic force microscope image of double and quadruple wedge structures.} (\textbf{a}) Scanning electron micrographs of cross sections for single wedges. The left column (${\theta}=$8, 9.5, 11, 14$^{\circ}$) shows cross sections of oxide wedges processed without HMDS, and the right column (${\theta}=$18.5, 23.5, 27, 55$^{\circ}$) shows the structure processed with HMDS. Within each of these cases, the wedge angle can be accurately controlled by adjustment of photoresist thickness. (\textbf{b}) Atomic force microscope image of double wedge and quadruple wedge (top and bottom surfaces are superimposed).}
\label{Method}
\end{centering}
\end{figure}
Hexamethyldisilazane (HMDS) was additionally deposited prior to photoresist coating for wedge angles larger than 20$^{\circ}$ (right column of Fig. 5a). For wedge angles in a range of 8$^{\circ}$$-$20$^{\circ}$, photoresist was spin-coated on the oxide layer without HMDS (left column of Fig. 5a). Photoresist thickness was used to adjust the wedge angle. The photoresist pattern acted as an etch mask during immersion in a buffered hydrofluoric acid (HF) solution, and the resist was removed once wet etching was completed. The second  SiO$_{2}$ layer was grown at the same temperature as the etched wedge disk, and the thickness of the oxide was accurately controlled via duration of  thermal oxidation. Again, the photoresist was patterned on the oxide layer and the wet etch proceeded to create additional wedge angles. After cleaning to remove photoresist and residues, XeF$_{2}$ dry etch was applied to etch the Si.   \\
\begin{figure}[h]
  \begin{centering}
  \includegraphics[width=8cm]{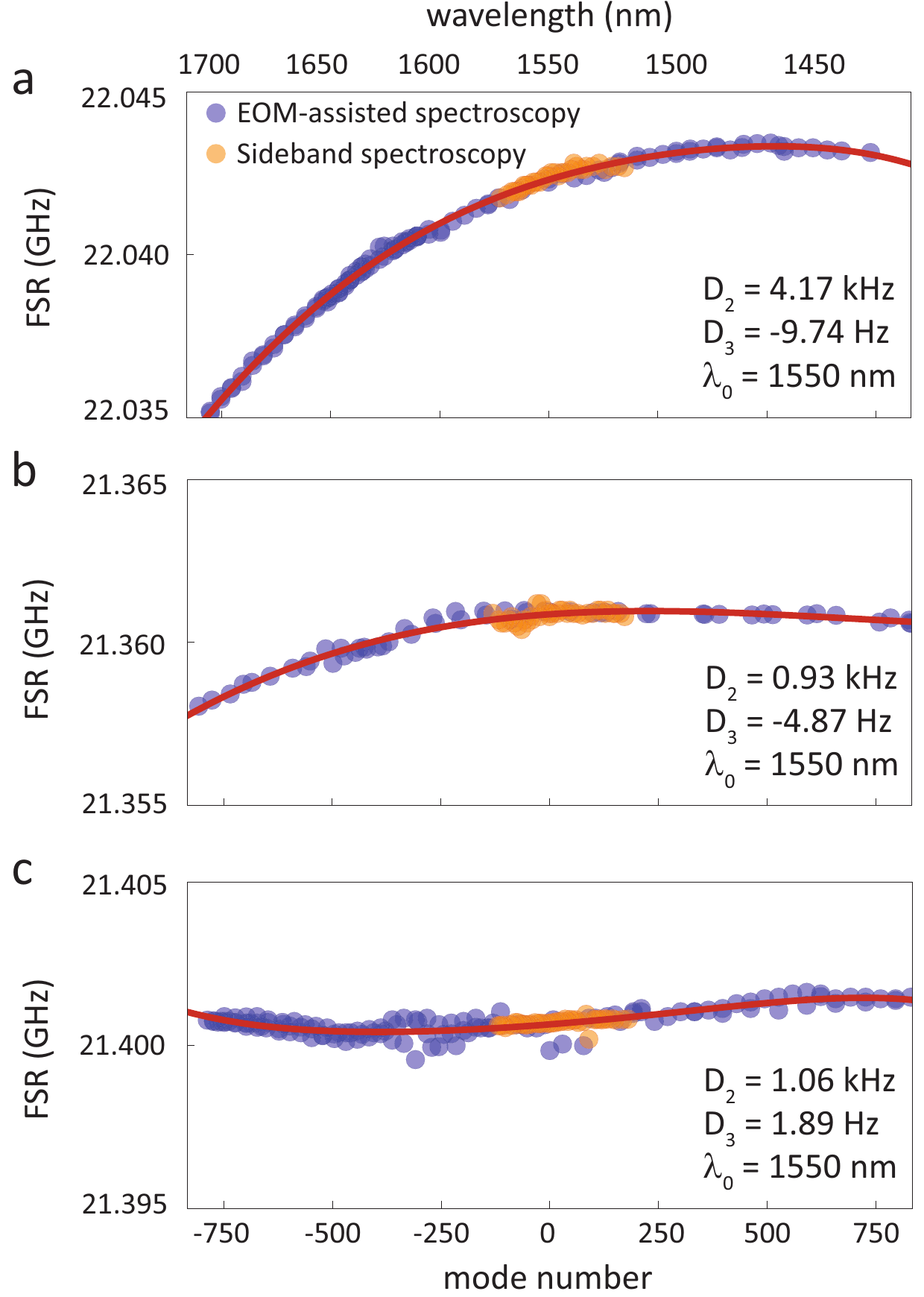}
   \captionsetup{singlelinecheck=no, justification = RaggedRight}
    \caption{\textbf{Dispersion parameters fitting} (\textbf{a}) Measured FSR of single-wedge disk and the Taylor series fit (solid line). Blue dots represent the measurement value using the EOM comb method (Fig. 3a), and red dots are measured by sideband spectroscopy\cite{JiangLi_OE,Pascal_2014_PRL} for comparison. The FSR is 22.042 GHz at 1550 nm, and dispersion parameters are determined as the coefficients of Taylor series at the same wavelength. (\textbf{b-c}) Same as panel \textbf{a}, but for quadruple-wedge disks. FSR: 21.36 GHz for (\textbf{b}), 21.40 GHz for (\textbf{c}) at 1550 nm. }
\label{Method}
\end{centering}
\end{figure}
\indent
\\
\textbf{Finite element simulation.} The numerical calculations were implemented using the COMSOL multiphysics package. Fibre parameters used in the Fig.1 were from \mbox{ref [\!\!\citenum{White_1979,Miya_1979_SC,QC_design_1984}]}, and the values for fused silica refractive index were taken from the Sellmeir equation of \mbox{ref [\!\!\citenum{Sellmeier}]}. Resonator dispersion was calculated via an iterative approach\cite{Pascal_2009NP}. Fig. 1a-b use a resonator thickness of 8 ${\mu}m$, and Fig. 1c-d use outer wedge heights of 3 ${\mu}m$ and 5 ${\mu}m$ for the angle ordering 45$^{\circ}$$-$10$^{\circ}$ (red) and 10$^{\circ}$$-$45$^{\circ}$ (blue), respectively. Fig. 4 uses a resonator geometry obtained from fabricated devices. Oxide thickness was measured using Filmetrics model F40 and Dektak 3ST profilometer, and the wedge profile was obtained using an atomic force microscope (Bruker Dimension ICON) and scanning electron microscope (Hitachi S-4100). Fig. 5b shows an atomic force microscope image of top and bottom surfaces of double-wedge and quadruple-wedge disks. The top surface profile was obtained from the disk before silicon dry etch (5th step in Fig. 2a), and then the silica disk was completely removed by BHF and the silicon surface morphology was scanned for the bottom surface profile. \\
\\
\textbf{Dispersion parameters fitting.} The resonance frequencies of one mode family can be described as a Taylor series shown in eqn.(2), and the coefficients of the series correspond to dispersion parameters at $\mu=0$. The FSR can also be expressed as a Taylor series with the same coefficients in eqn.(2). 
\begin{eqnarray}
FSR (\mu) ={D}_{1}+D_{2}\cdot {\mu}+\frac{D_{3}}{2!}{\mu}^{2}+\cdots
\end{eqnarray}
Fig. 6 shows the measured FSR of single-wedge and quadruple wedge structures (Fig.4), and the fitted polynomial curves (eqn.(3)) to the experimental results. Here, FSR was measured using both EOM-assisted (Fig. 3a, 300-nm-bandwidth) and sideband\cite{JiangLi_OE} (70-nm-bandwidth) spectroscopy methods. The variation of FSR is very small ($<$10 kHz /mode), thus it is necessary to accumulate the variation of FSR throughout multiple FSR\cite{JiangLi_OE} in order to fulfill the condition that $\mid{D}_{2}\mid \cdot {\mu}> {\Delta}f, \mid\frac{D_{3}}{2}\mid\cdot{\mu}^{2}> {\Delta}f$ where ${\Delta}f$ is the resolution in determination of the FSR (approximately 100 kHz).  \\
%%%%%%%%%%%%%%%%%%%%%%%%%%%%%%%%%%%%%%%%%%%%%%%%%%%%%%%%%%%%%%%%%%%%%%%%%%%%
%%%%%%%%%%%%%%% Acknowledgments, Figures, and References Section %%%%%%%%%%%%%%
%%%%%%%%%%%%%%%%%%%%%%%%%%%%%%%%%%%%%%%%%%%%%%%%%%%%%%%%%%%%%%%%%%%%%%%%%%%%

\hspace{10 mm}

\textbf{Acknowledgments} We gratefully acknowledge support from the Defense Advanced Research Projects Agency under the QuASAR program, the National Institute of Standards and Technology, the Kavli Nanoscience Institute and the Institute for Quantum Information and Matter, an NSF Physics Frontiers Center with support of the Gordon and Betty Moore Foundation. 
\hspace{10 mm}

\textbf{Author Information} Correspondence and requests for materials should be addressed to K.J.V. (vahala@caltech.edu).
\hspace{10 mm}

\textbf{Competing financial interests} The authors declare no competing financial interests. 
\hspace{10 mm}

\bibliography{ref_v7}

\end{document}